\documentclass[11pt]{article}
\usepackage{moriond,epsfig}

\begin{document}
\vspace*{4cm}
\title{METAL ENRICHMENT OF THE INTRA-CLUSTER MEDIUM: RAM-PRESSURE STRIPPING OF 
CLUSTER GALAXIES}

\author{W. DOMAINKO$^1$, W. KAPFERER$^1$, S. SCHINDLER$^1$, E. v. KAMPEN$^1$,
S. KIMESWENGER$^1$, \\ M. MAIR$^1$ \& M. RUFFERT$^2$}

\address{$^1$Institut f\"ur Astrophysik,Leopold-Franzens-Universit\"at 
Innsbruck\\ 
Techniker Stra\ss e 25, A-6020 Innsbruck Austria \\
http://astro.uibk.ac.at/astroneu/hydroskiteam/index.htm\\
$^2$Department for Mathematics and Statistics, University of Edinburgh\\
JCMB, Mayfield Road, Edinburgh, EH9 3JZ, UK}

\maketitle\abstract{
We present numerical simulations of the dynamical and chemical evolution of 
galaxy clusters. X-ray spectra show that the intra-cluster medium contains
a significant amount of metals. As heavy elements are produced in the stars
of galaxies material from the galaxies must have been expelled to enrich the 
ambient medium. We have performed hydrodynamic simulations investigating 
various processes. In this presentation we show the feedback from gas which 
is stripped from galaxies by ram-pressure stripping. The efficiency, 
resulting spatial distribution of the metals and the time dependency of this 
enrichment process on galaxy cluster scale is shown.}

\section{Introduction}
Clusters of galaxies contain a hot and thin plasma inbetween the cluster 
galaxies -- the Intra-Cluster Medium (ICM) which can be observed in X-rays.
X-ray spectra of the ICM show metal lines (Fukazawa et. al. 1998)
\cite{fukazawa} 
which indicates that the gas is not only of primordial origin as metals are 
generally produced in stars. The
amount of metals in the ICM is of the same order of magnitude as the amount
of metals found in the cluster galaxies. This means that processed material  
from stars and supernovae must have been ejected into the ICM. Possible 
enrichment mechanisms of
the ICM are the feedback from Intra-Cluster Supernovae (Domainko et al. 2004)
\cite{domainko} and material expelled by cluster galaxies. Processes which
can remove gas from cluster galaxies are ram-pressure stripping (Gunn \& Gott
1972)\cite{gunngott}, galactic winds (De Young 1978)\cite{deyoung}, 
galaxy-galaxy
interactions and jets from active galaxies. Two of these processes 
(ram-pressure stripping, galaxy-galaxy interactions) are 
triggered by their surroundings whereas the other two processes 
(galactic winds and jets) are triggered
by violent internal processes. The efficiency, time dependences
and spatial distributions of these mechanisms are poorly understood. In this 
paper we investigate the effect of ram-pressure stripping on the chemical
evolution of the ICM.

\section{Numerical Method}
We use combined N-body and Hydrodynamical simulations to compute the effect
of different enrichment processes on the ICM. The simulations are performed 
on galaxy cluster scale to investigate the efficiency, time dependence and 
spatial distribution of the chemical evolution of the ICM. Large-scale
structure formation is derived with a N-body tree code with an additional
semi-numerical model for galaxy formation (van Kampen et al. 1999)
\cite{vankampen}.
On this background potential a shock capturing grid based PPM (Colella \&
Woodward 1984)\cite{collela} hydrodynamic simulation is performed. 
The effect of 
ram-pressure stripping according to the local properties of the ICM and
the properties of the galaxies as well as the 
effect of galactic winds (see Kapferer et al. this volume)\cite{kapferer} are 
included in
the calculations. The hydrodynamic simulation is obtained on four nested grids
(Ruffert 1992)\cite{ruffert}. This technique allows to cover the cluster center
where most of the stripping is expected to happen with high resolution 
(cell size comparable to galaxy size) and also
to investigate the effect of gas rich galaxies falling in towards the cluster
center (in e.g. hierarchical merger events). Metallicity is used as a tracer
to follow the enriched material.

\section{Ram-pressure stripping}
Galaxies moving fast through an ambient medium (in galaxy clusters the ICM) 
can suffer from environmental 
interactions. In particular the Inter-Stellar Medium (ISM) of the galaxy is 
effected by the ram-pressure of a surrounding medium when its host galaxy
moves with a sufficiently high velocity. Additionally the value of the 
ram-pressure also depends on the properties (density) of the ICM
(the environmental influence will increase with increasing density). 
If the force due to the ram-pressure
exceeds the restoring gravitational acceleration the gas in this region
will be stripped away from the affected galaxy. This can influence both
the stripped galaxy and the stripping initiating ICM. On the one hand
a depletion of ISM leads to a decreased star formation rate which
may explain the Butcher-Oemler effect (Buther \& Oemler 1984) \cite{butcher} 
and the formation of S0 galaxies 
(Dressler et al. 1997)\cite{dressler}. 
On the other hand the stripped material will enrich the surrounding
medium with heavy elements. This is the effect we study with
our simulations.

\section{Dynamical State}
Mergers of galaxy clusters have strong effects on the physical quantities of
the ICM. The most prominent features of merger events are shock waves
which we also clearly see in our simulation (due to good shock resolution 
of our hydrodynamic treatment ). 
Shock waves represent a step in the density distribution of the cluster. 
A sudden increase in the surrounding density will also result in an 
increased stripping rate due to ram-pressure stripping in an affected 
cluster galaxy. Mergers
also change the density and temperature distribution of the ICM which then 
obviously influences environmental dependent enrichment mechanisms like
ram-pressure stripping. During and after merger events the ICM is mixed 
due to turbulent gas motions. This can result in a significant change of the
metallicity distribution in such systems. We ran several models with different
merger scenarios to investigate the above mentioned effects.

\begin{figure}[h]
\center
\psfig{figure=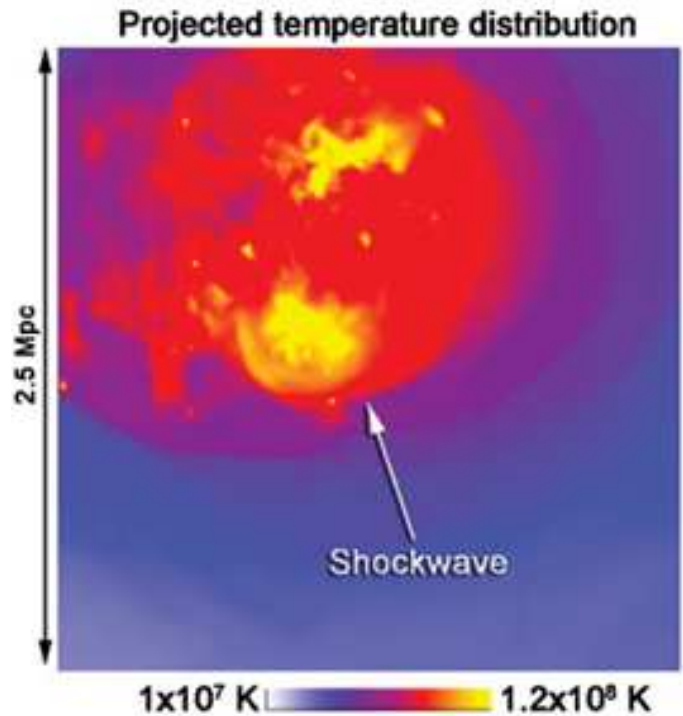,height=4in}
\caption{Post merger state of a simulated galaxy cluster. A huge shock
wave can clearly be seen.}
\end{figure} 

\section{First Results}
In first simulations we investigate the chemical evolution of the ICM
from redshift $z=1$ to $z=0$.
In the case of ram-pressure stripping we clearly see that the process is
getting more efficient when moving closer to the cluster center. This
can be understood in terms of higher velocities and higher ambient
densities strip galaxies more efficiently as discussed above.
We also study different merger events and their influence on 
the properties and the distribution of the enriched material in the ICM. 
In the central region of the cluster some mixing due to merger events
can be seen. For a better interpretation of cluster observations we model 
specific
clusters (e.g. Coma). From this simulations constraints on the merger 
history of the modeled systems can be made. Additionally we produce X-ray 
brightness maps,
temperature maps and metallicity maps which can directly be compared
to X-ray observations.

\begin{figure}[h]
\center
\psfig{figure=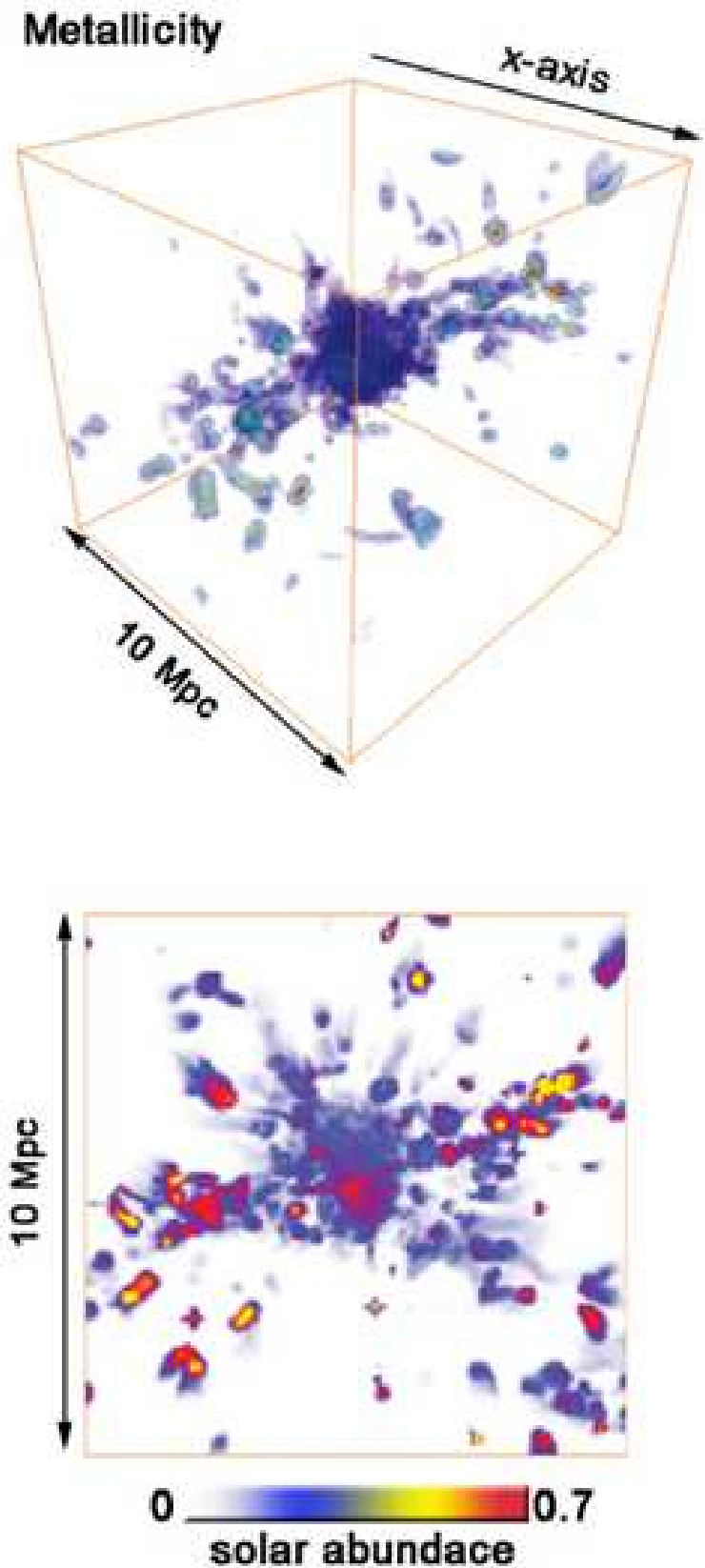,height=9in}
\caption{The distribution of stripped, enriched material
which is produced since redshift $z=1$. The lower image is a
projection along the X-axis. Some mixing at the cluster center can be seen.}
\end{figure} 

\section*{Acknoledgement}
The authors acknowledge the support of the European Commission through grant
number HPRI-CT-1999-00026 (the TRACS Program at EPCC) and the Austrian
Science Foundation (FWF) through grant number P15868.

\end{document}